\documentclass[superscriptaddress, amsmath, amssymb,  prd, showpacs, floatfix ,preprintnumbers, natbib, onecolumn ,longbibliography,tikz,border=10pt]{article}
\usepackage[utf8]{inputenc}
\usepackage[margin=1in]{geometry}  
\usepackage{graphicx}
\usepackage{amsmath}
\usepackage[table]{xcolor}  
\usepackage{colortbl}       
\usepackage{subcaption}
\usepackage{multicol}
\usepackage{multirow}
\usepackage{float}

\usepackage[affil-it]{authblk}

\usepackage{dcolumn}
\usepackage{bm}

\usepackage{hyperref}

\begin{document}

\title{Patient-Specific Modeling of Dose-Escalated Proton Beam Therapy for Locally Advanced Pancreatic Cancer}
\author[1]{Melissa Anne McIntyre}
\author[1,2]{Jessica Midson}
\author[3,2]{Puthenparampil Wilson}
\author[4,2]{Peter Gorayski}
\author[5]{Cheng-En Hsieh}
\author[6]{Shu-Wei Wu}
\author[1,7]{Eva Bezak}

\affil[1]{Allied Health \& Human Performance, University of South Australia, Adelaide, SA 5001, Australia.}
\affil[2]{Department of Radiation Oncology, Royal Adelaide Hospital, Adelaide, SA 5000, Australia.}
\affil[3]{UniSA STEM, University of South Australia, Adelaide, SA 5001, Australia.}
\affil[4]{Australian Bragg Centre for Proton Therapy and Research, Adelaide, SA 5000, Australia.}
\affil[5]{Departments of Medical Imaging and Radiological Sciences, Radiation Oncology, and Cancer Genome Research Center, Chang Gung Memorial Hospital at Linkou and Chang Gung University, Taoyuan, Taiwan.}
\affil[6]{Department of Radiation Oncology, Chang Gung Memorial Hospital at Linkou and Chang Gung University, Taoyuan, Taiwan.}
\affil[7]{Department of Physics, University of Adelaide, Adelaide, SA 5005, Australia.} 

\maketitle

\begin{abstract}
	\textbf{Purpose:} This study explores the feasibility of dose-escalated proton beam therapy (dPBT) for Locally Advanced Pancreatic Cancer (LAPC) patients by modeling common patient scenarios using current clinically-adopted practices. \par
    \textbf{Methods:} Five patient datasets were used as simulation phantoms, each with six tumour sizes, to systematically simulate treatment scenarios typical in LAPC patients. Using the Raystation treatment planning system, robustly-optimised dPBT and stereotactic ablative radiotherapy (SABR) treatment plans were created with a 5\,mm margin allowing for intra- and inter-fraction anatomical changes. following clinically-adopted protocols. Safe dose-escalation feasibility is assessed with dose metrics, tumour control (TCP) and normal tissue complication probabilities (NTCP) for average and worst-case intra-fraction motion scenarios. Significance testing was performed using a paired student's t-test.\par
    \textbf{Results:} Dose-escalation feasibility is largely dependent on tumour size and proximity to critical structures. Minimal therapeutic benefit was observed for patients with $>4.5$\,cm tumours, however for tumours $\leq4.5$\,cm dPBT TCPs of 45-90\% compared to SABR TCPs of 10-40\% ($p<0.05$). The worst-case scenario dPBT TCP was comparable to SABR. Hypofractioned dPBT further improved this result to $>90$\% ($p<0.05$) for tumours $\leq4.5$\,cm.\par 
    \textbf{Conclusion:} Safe dPBT is feasible for patients with targets up to the median size and see a significant therapeutic benefit compared to the current standard of care in SABR. A patient-specific approach should be taken based on tumour size and surrounding anatomy.
\end{abstract}

\maketitle


\section{Introduction}
Locally advanced pancreatic cancer (LAPC) has one of the poorest prognoses among common cancers. It is characterised by late presentation of symptoms, often where the cancer has advanced beyond the point of surgical resection at the time of diagnosis, as well as high recurrence rates, morbidity and mortality \cite{cancers13061277, ausCancerData}. Despite significant advances in therapeutic methods, the 5-year survival rate of LAPC is currently 13\% in Australia \cite{ausCancerData}, and similar trends worldwide, suggesting unmet clinical demand for effective treatments.\par 
Current clinical approaches, such as targeted tumour-sensitising therapies, or radiation therapy \cite{cancers12010163}, are severely constrained by the intra-tumour and surrounding environment. Notably, high levels of hypoxia within the tumour environment reduces the efficacy of radiation \cite{hypoxPC}. Additionally, the close proximity of critical gastrointestinal (GI) organs limits the radiation dose deliverable to the tumour, and the presence of a blood barrier, reduces the permeability of therapeutic or sensitising agents \cite{stromPC}.\par
Proton therapy (PBT) holds promise in the treatment of LAPC due to its ability to provide superior tissue sparing compared to photon-based therapies, such as stereotactic ablative radiation therapy (SABR) \cite{hitchcock_PBT}, potentially allowing for dose escalation while minimising toxicity. Although SABR is a current standard for radiation therapy of LAPC, delivering highly conformal, ablative doses to tumours, its effectiveness is limited by the hypoxic tumour environment and proximity to gastrointestinal (GI) organs, which restricts the high doses necessary for optimal tumour control \cite{cancers12010163,LIN201955,Crane2016_doseEsc, eckstein2023_pbt4pc}. The advantages of PBT, including integral dose sparing and potentially enhanced biological effects, offer an opportunity to safely escalate doses and achieve sufficient tumour control with reduced GI toxicity probability. Evidence from previous studies supports PBT’s tolerability in LAPC patients, with some observed improvements in local control \cite{hitchcock_PBT, hiroshima2019, JGO64612}. For instance, in a study with 15 unresectable LAPC patients prescribed 59.4\,GyRBE of PBT, a median survival of 24 months was observed \cite{hitchcock_PBT}. Similarly, 42 patients receiving high dose PBT achieved overall survival times of 13.1, 28.4, and 42.5 months for doses of 50, 54–60, and 67.5\,GyRBE, respectively, compared to photon therapies with a median survival of 15.7 months at 50-50.4\,Gy \cite{hiroshima2019}. These findings, along with evidence that PBT may provide improved local control compared to photon-based therapies, suggest that dose escalation is essential to enhance overall survival and local control. However, a careful assessment of risks to the surrounding anatomy remains critical.\par
This study does not seek to compare photon and proton beam therapies for LAPC, but rather explores the feasibility of PBT achieving dose escalation to improve tumour control while maintaining risk of toxicity compared to a current standard of care in SABR. Although prior research has explored dose-escalated PBT outcomes for LAPC \cite{hiroshima2019,bouchard2009_doseEsc}, this study is unique in modeling tumour control and toxicity by systematically varying patient characteristics and radiobiological sensitivities to understand their relative impact. The SABR and PBT plans were created using clinically utilised protocols, ensuring their relevance to current clinical practice. This approach establishes both a plausible clinical range for these scenarios and a critical “failure” point where radiation treatments become ineffective. Moreover, the study is novel in the potential of comparison with real long-term patient follow-up data as currently implemented clinical protocols are followed to create the treatment plans, as opposed to theoretical protocols. By simulating both average and worst-case inter- and intra-fraction anatomical shifts, we generate outcome probabilities in the best and worst-case scenarios. This allows us to estimate, within data limitations, the tumour control and normal tissue complication risks under typical treatment conditions often encountered in LAPC radiation treatments.\par

\section{Materials \& methods}
\subsection{Patient datasets}
Five de-identified retrospective patient datasets (three male, two female) were acquired from the Royal Adelaide Hospital with appropriate ethics approval (SA Health HREC no. 18587, UniSA ethics no. 206055), and with all clinical targets and organs at risk (OARs) segmented. The clinical target was located in the head of the pancreas for three patients, and in the body/tail for the remaining two, thus reflecting incidence statistics where head cases are more common than body/tail cases \cite{sun2023_lapctargetsizes, fortner1996_lapctargetsizes}. The relative positions of the organs surrounding the pancreas, namely the stomach, bowel and duodenum did not vary significantly in the treatment region between patients. Further, a wide range of patient sizes were not needed for this study as most LAPC patients present with weight-loss at diagnosis \cite{lapcweightloss} and thus larger patients are uncommon. As such, five datasets were ample to reflect variations between patients.

\subsection{Simulation of treatment scenarios}
On each patient, the clinical target (internal target volume, ITV) was expanded/contracted such that its longest axial dimensions measured at 2.5, 3.5, 4.5, 5.5, 6.5 and 7.5\,cm, previously reported as a common range of clinical target sizes for LAPC \cite{sun2023_lapctargetsizes, fortner1996_lapctargetsizes}, totaling 30 treatment scenarios (five patients $\times$ six targets).\par
A large number of unique CT datasets was not required for this study as each patient was used as a ``simulation phantom" where clinical targets can be drawn with various sizes, positions and Hounsfield Units in the treatment planning system. This technique has been previously adopted in the literature \cite{bouchard2009_doseEsc, cancers12092578} and was used here such that a typical clinical range of tumour characteristics could be simulated with other influencing factors being systematically controlled to understand their impact.\par

\subsection{Treatment planning approach}
\subsubsection{Planning parameters \& objectives}
For each simulated patient scenario, a SABR and dose-escalated pencil beam scanning PBT (dPBT) plan was prepared according to the criteria outlined in two clinically utilised protocols, the SABR MASTERPLAN \cite{masterplan} and dPBT protocol from the Chang Gung Memorial Hospital, using the \textsc{RayStation} treatment planning system (v2023B, RaySearch Laboratories, Sweden). Dose calculations were performed using RayStation's Monte Carlo algorithm (v5.6) on a grid with $0.2\times0.2\times0.2\,$cm$^3$ voxels. The SABR plans were developed using a Varian Truebeam STx beam model (beam energy 6\,MV), whilst the dPBT plans used an IBA ProteusOne beam model (relative biological effectiveness = 1.1, beam energies 140-200\,MeV, spot size $\sigma$ = 4-5\,mm and hexagonal spot arrangement with spot spacing 6-7.3\,mm) with generic CT number tables. Doses of 40\,Gy in 5 fractions (biologically effective dose (BED) 72\,Gy) for SABR and 66\,GyRBE in 22 fractions (BED 86\,GyRBE) for dPBT were prescribed. The duodenum, stomach and small bowel were considered dose-limiting structures and thus dose coverage to the ITV was compromised to ensure they remained below tolerance, as per the respective protocols. The coverage and organ tolerance criteria for each modality are summarised in Table \ref{tab:clinical_goals}.\par 

\subsubsection{Intra- \& inter-fraction motion management}
The internal target volume (ITV) was segmented to encompass the gross target volume (GTV) position at both full inhalation and exhalation \cite{masterplan}. A planning target volume (PTV) and planning risk volume (PRV) approach was employed to ensure robustness in SABR plans, with a 5\,mm expansion of the ITV (PTV = ITV + 5\,mm) and of the OARs (PRV = OAR + 5\,mm) to keep all dose-limiting structures within tolerance limits. The dPBT plans were robustly-optimised using a $\pm$5\,mm setup uncertainty and a $\pm$3.5\% proton range uncertainty (lateral displacement), resulting in a total of 21 scenarios. The ITV and dose-limiting structures were included in the robust scenario simulation where each region of interest was shifted $\pm$5\,mm in each patient plane.\par 
The optimisation process ensured that all OAR clinical goals were met even in the worst-case anatomical arrangement. The 5\,mm margin was assumed to incorporate patient setup, intra- and inter-fraction anatomical shifts that occur throughout treatment for both modalities. This margin has been previously recommended in the published literature where these shifts were quantified \cite{5mmmargin1}, and is recommended in the Masterplan protocol as the conservative margin to use \cite{masterplan}.\par
In both protocols used in this work, patients are CBCT imaged daily to assess inter-fraction anatomical changes. In the event that an anatomical change occurs outside the defined 5\,mm margin, treatment would not proceed and would be delayed until the following day, where they are imaged and reassessed. Pre-treatment preparations also require the patient to fast prior to treatment, and that the treatment fraction always be delivered at the same time of day to minimise variation in bowel gas and resultant organ shifts.

\subsubsection{Beam arrangement}
\begin{figure}[ht]
   \begin{center}
   \includegraphics[width=10cm]{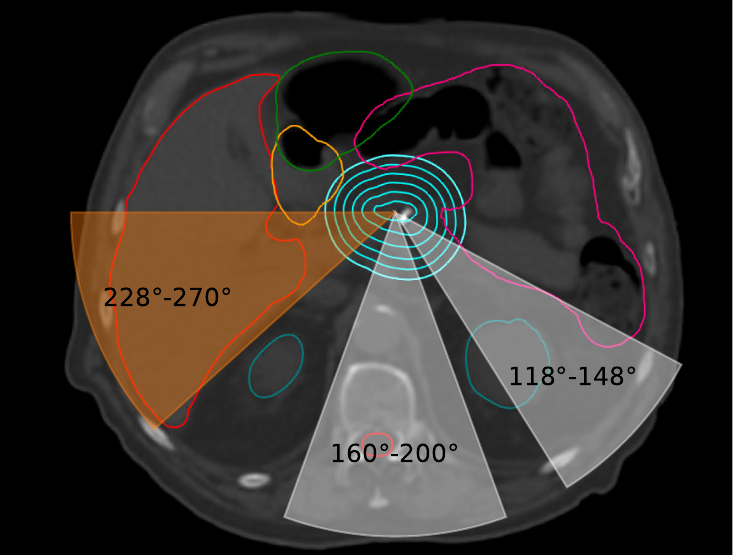}
   \caption{Range of dPBT beam configurations for all dPBT plans demonstrated on one patient simulation phantom. Gantry angles indicated in white are coplanar and in orange are non-coplanar.
   \label{fig:beamAnglePlot} 
    }  
    \end{center}
\end{figure}

Two full beam arcs were used for the SABR plans with small degrees of target-OAR overlap, whilst a third arc was utilised for high overlap cases. Similarly, two to three beams were chosen for dPBT depending on the degree of OAR overlap. 
Two or three full beam arcs and static beams for SABR and dPBT, respectively, were used for each target volume scenario depending on the complexity of overlap between the ITV and OARs. The dPBT gantry angles were chosen independently for each treatment scenario based on patient anatomy. Beams were limited to the posterior and lateral directions to mitigate the impact of range uncertainties for beams directed through highly variable anatomy in the anterior region, as shown in Figure \ref{fig:beamAnglePlot} \cite{AMOS2022e188}. In scenarios where it was not feasible to avoid both kidneys with beam placement, a beam would only be placed through one with the other being spared or receiving minimal dose. Beam paths through in-homogeneous tissue were avoided, such as the spinal cord (where possible) and the couch edge (in every scenario).\par

\subsubsection{Reliability \& validity}
To ensure all treatment plans were to a clinical standard and feasibly deliverable, they were independently reviewed and approved by a radiation therapist (J.M.), two radiation oncologists (P.G., R.C.E.H.) and a medical physicist (S.W.W.).

\begin{table}[htbp]
\centering
\caption{Primary dosimetric objectives for SABR and dPBT treatment plans obtained from their respective protocols.}
\label{tab:clinical_goals}
\renewcommand{\arraystretch}{1.5} 
\begin{tabular}{|l|c|c|c|c|}
\hline
\multirow{2}{*}{Volume} & \multicolumn{2}{c|}{SABR} & \multicolumn{2}{c|}{dPBT} \\
\cline{2-5}
                        & Metric          & Objective (Gy$^\S$)   & Metric          & Objective (GyRBE$^\S$)  \\
\hline
ITV                     & D$_{\text{99\%}}$  & $>$33.0              & D$_{\text{90\%}}^{\ddagger}$ & $\geq$62.7   \\
                        & D$_{\textsubscript{max}}^{\ast,\ddagger}$  & 44.0--52.0       & D$_{\textsubscript{max}}^{\ast\ast,\ddagger}$ & $\leq$75.9 \\
\hline
PTV                     & D$_{99\%}$        & $>$30.0              &                  &                 \\
                        & D$_{90\%}^{\dagger}$ & $>$40.0            &                  &                 \\
\hline
Duodenum, Stomach \& Bowel (PRV) & V$_{30\%}$  & $<$5\,cm$^3$ & V$_{\textsuperscript{40GyRBE}}$  & 5.0\,cm$^3$  \\
                        & V$_{20\%}$  & $<$20\,cm$^3$  & D$_{\textsubscript{max}}^{\ast\ast}$ & 46.2 \\
                        & D$_{\textsubscript{max}}^{\ast}$ & $<$33.0 ($<$38.0) &  &  \\
\hline
Cord                    & D$_{\textsubscript{max}}^{\ast}$ & $<$20.0  & D$_{\textsubscript{max}}^{\ast\ast}$ & 39.0 \\
\hline
Kidneys L/R             & V$_{\textsuperscript{10Gy}}$ & $<$10\%   & V$_{\textsuperscript{18Gy}}$ & $<$33\% \\
Kidneys Combined        & V$_{\textsuperscript{12Gy}}$ & $<$40\%   & & \\
\hline
Liver                   & V$_{\textsuperscript{12Gy}}$ & $<$40\%   & V$_{\scriptscriptstyle < 1\,\mathrm{Gy}}$  & $>$400.0\,cm$^3$ \\
\hline
Spleen                  & D$_{\textsubscript{mean}}$ & $<$4.5     & D$_{\textsubscript{mean}}$ & $<$4.5 \\
\hline
Skin                    & D$_{\textsubscript{max}}^{\ast}$ & $<$25.0  & & \\
\hline
\end{tabular}
\vspace{1ex} 
\small
$^\S$unless indicated otherwise, $^\dagger$non-OAR overlapping target, $^\ast$d0.5\,cm$^3$, $^\ddagger$must fall within ITV, not PTV or non-nominal robust scenario, $^{\ast\ast}$d0.03\,cm$^3$
\end{table}

\subsection{Plan evaluation}
A robust evaluation was performed for all plans and modalities using a patient shift margin of $\pm5\,$mm and a $\pm3.5\%$ proton range uncertainty (dPBT only). The clinical goals for all dose-limiting structures were met under the worst case and nominal anatomy position scenario for each plan. In this paper, worst-case anatomy placement refers to the positions of the surrounding dose-limiting structures and tumour with the highest degree of overlap or closest proximity. The cumulative dose volume histograms (cDVHs) in all robustness scenarios and regions of interest were exported from Raystation for analysis.

\subsection{Evaluating TCP}
The TCP was computed using a model based on Poisson statistics for the ITV to estimate local tumour control \cite{tcppoisson}. The ITV was selected for this computation as it is assumed to contain the primary tumour and surrounding micro-disease. The ITV differential DVH (dDVH), made up on $N$ bins \{$D_i, v_i$\}$_{i=0}^{N}$, was exported from Raystation and the ITV was divided into smaller fractional sub-volumes, $v_j$, equal to the size of a single dose-grid voxel (0.2$\times$0.2$\times$0.2\,cm$^3$). To account for variations of radiosensitivity across the target volume, each sub-volume $v_j$ was assigned an intrinsic radiosensitivity $(\alpha_j, \beta_j)$, uniformly sampled from linear-quadratic model parameters derived from fits to experimental \textit{in vitro} clonogenic cell survival data of LAPC cell lines based on published literature (Table \ref{tab:alphabeta_ratio}). 
For studies where $\alpha$/$\beta$ ratios were reported in the presence of radiosensitisers, only the control (radiation only) $\alpha$/$\beta$ ratios were extracted for use in this study. To our knowledge, $\alpha$/$\beta$ ratios for PC cell lines have not been quantified under proton irradiation. The impact of this limitation is further addressed in the Discussion section. For each sub-volume $v_j$ corresponding to dose bin $D_i$, the TCP is calculated as:
\begin{eqnarray}
    TCP(\alpha_j, \beta_j)_i &=& \prod_{j} \exp\bigg[-\rho\cdot v_j\cdot S(D_i)_j\bigg]
\end{eqnarray}
here $\rho$ is the density of clonogens, assumed to be $10^8$\,cells/cm$^3$ as a conservative value from the plausible range in published literature \cite{clonogendensityrange}, and $S(D_i)_j$ is the probability of cell survival in the sub-volume. The cell survival probability of sub-volume $v_j$ is defined as follows:
\begin{eqnarray}
    S(D_i)_j &=& \exp\bigg[-n\cdot d_i\cdot(\alpha_j+\beta_j\cdot d_i)\bigg]
\end{eqnarray}
where $d_i$ is the dose per fraction and $n$ is the total number of fractions in the treatment. The TCP of the whole tumour is represented by summing the TCP of each sub-volume weighted by the fractional volume contained in dose bin $v_i$:
\begin{eqnarray}
    TCP &=& \sum_{i} v_i\cdot TCP(\alpha_j, \beta_j)_i
\end{eqnarray}
where $\sum_i v_i = 1$. The effects of accelerated re-population during the course of treatment were considered negligible in this work due to the large kick-off time of 49 days relative to the total treatment time of five weeks for dPBT and two weeks for SABR \cite{Gholami2022_Kickofftime}. Additionally, the effects of hypoxia were omitted from this work and will be subject to future study.

\begin{table}[htbp]
\begin{center}
\caption{Linear-quadratic model fit parameters $\alpha$, $\beta$ derived from clonogenic cell survival studies.
\label{tab:alphabeta_ratio}
\vspace*{2ex}
}
\begin{tabular} {|l|l|c|c|c|}
\hline
Study & Cell Line & $\alpha$ & $\beta$& $\alpha/\beta$\\
\hline
 Schwarz et al. \cite{schwarz2020_alphabeta} & MiaPaCa-2 & 0.36 & 0.025 & 14.4\\
 Shafie et al. \cite{shafie2013_alphabeta} & BxPC-2 & 0.28 & 0.019 & 14.7\\
Shafie et al. \cite{shafie2013_alphabeta} & PANC-1 & 0.29 & 0.021 & 13.8\\
Waissi et al. \cite{waissi2021_alphabeta} & MiaPaCa-2 & 0.45 & 0.025 & 18.0\\
Waissi et al. \cite{waissi2021_alphabeta} & PANC-1 & 0.27 & 0.024 & 11.2\\
Waissi et al. \cite{waissi2021_alphabeta} & BxPC-3 & 0.27 & 0.013 & 20.8\\
\hline
\end{tabular}
\end{center}
\end{table}

\subsection{Evaluating NTCP}
The differential DVHs for OARs of interest were exported from Raystation and converted to an equivalent dose under a standard fractionation regimen of 2\,Gy per fraction (EQD2) \cite{eqd2article}, assuming a standard normal tissue $\alpha/\beta$ of 3.0\,Gy. The NTCP was computed using the Lyman-Kutcher-Burman (LKB) model \cite{lyman1985_lkbmodel} with the Lyman-Wolbarst DVH reduction method \cite{lymandvhreduction}. This method converts the DVH of a non-uniformly irradiated volume into that of an equivalent, uniformly irradiated volume. This equivalent uniform dose (EUD) for a differential DVH with $N$ bins \{$D_i, v_i$\}$_{i=0}^{N}$ is expressed as:
\begin{equation}
    EUD = \bigg(\sum_i v_iD_i^{\frac{1}{n}}\bigg)^n\quad.
\end{equation}
To compute the NTCP as a function of EUD, we apply the following formulae:
\begin{equation}
    NTCP = \frac{1}{2\pi}\int_{-\infty}^{t}e^{-\frac{x^2}{2}}\text{dx} \quad,
\end{equation}
where
\begin{equation}
    t = \frac{EUD-TD_{50}}{mTD_{50}}\quad.
\end{equation}
$TD_{50}$ is the dose [Gy/GyRBE] at which the probability of the toxicity occurring is 50\%, $m$ is the slope of the sigmoid NTCP function, and $n\in[0,1]$ is an organ-specific parameter that defines its seriality. The LKB model parameters used in this work were derived from clinical data for select toxicities and are summarised in Table \ref{tab:lkb_params}.

\begin{table}[htbp]
\begin{center}
\caption{LKB parameters derived from clinical studies. An $\alpha$/$\beta$ ratio of 3\,Gy was assumed for all OARs.
\label{tab:lkb_params}
\vspace*{2ex}}
\begin{tabular} {|l|c|c|c|c|c|}
\hline
Study & Organ & Toxicity & $TD_{50}$ [Gy] & $m$ & $n$ \\
\hline
Pan et al. \cite{pan2003_ntcp} & \multirow{3}{*}{Stomach} & GI bleeding & 62.0 & 0.30 & 0.07 \\
\multirow{2}{*}{Burman et al.  \cite{burman1991_ntcp}} &  & Ulceration/ & \multirow{2}{*}{65.0} & \multirow{2}{*}{0.14} & \multirow{2}{*}{0.15} \\
&&Perforation&&&\\
\hline
\multirow{2}{*}{Burman et al.  \cite{burman1991_ntcp}} & \multirow{3}{*}{Small Bowel} & Obstruction/ & \multirow{2}{*}{55.0} & \multirow{2}{*}{0.16} & \multirow{2}{*}{0.15} \\
&&Perforation&&&\\
Reinartz et al.  \cite{reinartz2021_ntcp} & & Diarrhoea & 55.0 & 0.15 & 0.79 \\
\hline
Pan et al. \cite{pan2003_ntcp} & \multirow{2}{*}{Duodenum} & GI bleeding & 180.0 & 0.49 & 0.12 \\
Burman et al.  \cite{burman1991_ntcp} &  & Grade $\geq$3 toxicity & 299.1 & 0.51 & 0.19 \\
\hline
\end{tabular}
\end{center}
\end{table}

\subsection{Data analysis}
All analysis was performed using \textsc{Python} (v3.11) \cite{python3}, making ample use of packages \textsc{SciPy} \cite{Scipy2020} and \textsc{NumPy} \cite{numpy}. The dosimetric results and scripts used in this work are available online \cite{zenodolink}. A two-tailed Student's paired t-test ($\alpha=0.05$, significance at $p<0.05$) was used to determine whether the D$_{95\%}$, BED, TCP and NTCP differed between the two modalities in matched nominal and worst-case anatomical shift scenarios. In doing so, we determine in which target volume scenarios dose-escalation was achieved, through analysing physical dose and BED, and whether a statistically significant increase in TCP was achieved with dPBT without significantly worsened NTCP for select toxicities. All statistical analysis was designed after consultation with a biostatistician from the \textsc{University of South Australia}.\par 
In the following discussion, metrics for SABR and dPBT will be referred to as [metric](SABR) and [metric](dPBT), respectively. For example, $D_{95\%}$(SABR) is the dose to 95\% of a volume for the SABR modality. The volume of a region of interest (ROI) is henceforth expressed as V$_{\text{ROI}}$. The dose and treatment outcome metrics are analyzed as a function of V$_{\text{ITV}}$. To analyze how the dose metrics, TCP and NTCP behave with increasing target volume and thus proximity to GI organs, we utilise the metrics outlined in Table \ref{tab:overlapMetricDefs}.
\begin{table}[htbp]
\begin{center}
\caption{Metrics used to quantify the degree of ITV proximity to the stomach, bowel, and duodenum.}
\label{tab:overlapMetricDefs}
\vspace*{1ex}
\begin{tabular}{|c|p{9cm}|} 
\hline
\textbf{Expression} & \textbf{Description}\\
\hline
\multirow{3}{*}{
    $\begin{array}{rl}
    V_{\text{PRV}} = & V(\text{Duodenum}+5\,\text{mm}) \\
    & \quad+\, V(\text{Bowel}+5\,\text{mm}) \\
    & \quad\quad+\, V(\text{Stomach}+5\,\text{mm})
    \end{array}$ 
} & 
\multirow{3}{9cm}{Total volume of dose-limiting structures with a 5\,mm anatomical shift.} \\

& \\ & \\ \hline

\multirow{2}{*}{$V_{\text{PTV}} \cap V_{\text{PRV}}$} & 
Intersection volume between target and dose-limiting structures with a 5 mm anatomical shift. \\ \hline

\multirow{2}{*}{$\frac{V_{\text{PTV}} \cap V_{\text{PRV}}}{V_{\text{PTV}}}$} & 
Fraction of target volume occupied by dose-limiting structures within a 5 mm anatomical shift. \\ \hline
\end{tabular}
\end{center}
\end{table}

\subsection{Parameter sensitivity analysis}
The radiobiological parameters obtained from the published literature are subject to uncertainty and are based on \emph{in vitro} and/or photon-based studies. As we anticipate variations in these parameters for PBT  \cite{Mara2020_cellsurvivalstudy}, we performed a sensitivity analysis by varying all parameters in Tables \ref{tab:alphabeta_ratio} and \ref{tab:lkb_params} independently, and examined their impact on TCP and NTCP, respectively. For the TCP predictions, major sources of uncertainty arise in the choice of $\alpha/\beta$ and the clonogen density $\rho$. For the former, we varied $\beta$ such that $\alpha/\beta$ varied by 25\%, whilst the latter was varied by one order of magnitude $r=\pm1$ in $10^r$. The NTCP parameters $m$, $n$ and TD$_{50}$ were varied by $\pm$25\% for each toxicity considered in Table \ref{tab:lkb_params}.\par
The two protocols considered in this study utilise different fractionation regimens. Recall the study objective that we wish to increase the TCP using dPBT, with either improved or comparable NTCPs to SABR. Whilst maintaining the same total prescription dose for each respective protocol, we varied the dose per fraction of the dPBT plans until the dPBT NTCPs were comparable to, or no worse than, that of SABR. We then investigated the impact this would have on the TCP. 

\section{Results}
\subsection{Target coverage \& practicality of dose-escalation}
All OAR constraints were fulfilled in all robust scenarios for the dPBT plans, whilst the objectives of the SABR plans were met for the OARs and their associated PRVs as per Table \ref{tab:clinical_goals}. The dose-limiting nature of the proximal OARs resulted in target coverage deterioration, particularly for larger volumes in close proximity to GI organs.\par
\begin{figure}[htbp]
    \centering
    \begin{minipage}[b]{0.24\textwidth}
        \centering
        \includegraphics[width=\textwidth]{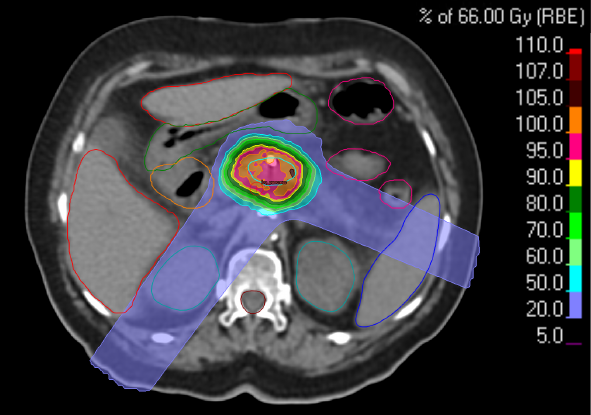}
        \subcaption{dPBT 3.5\,cm ITV}
    \end{minipage}%
    \begin{minipage}[b]{0.24\textwidth}
        \centering
        \includegraphics[width=\textwidth]{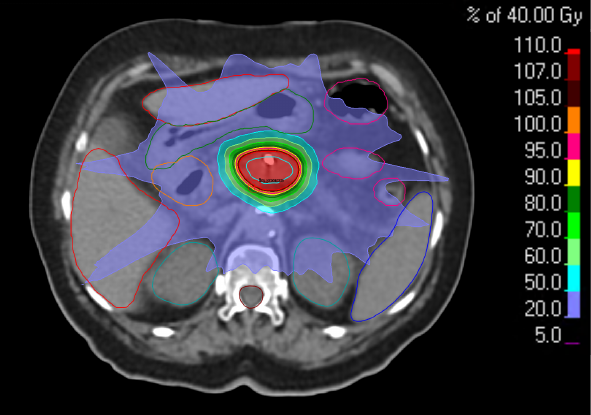}
        \subcaption{SABR 3.5\,cm ITV}
    \end{minipage}%
    \begin{minipage}[b]{0.24\textwidth}
        \centering
        \includegraphics[width=\textwidth]{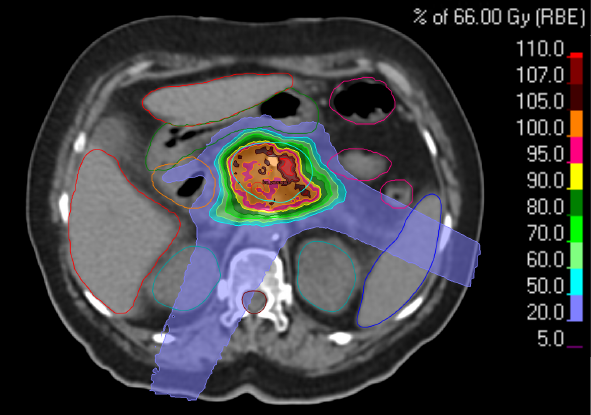}
        \subcaption{dPBT 5.5\,cm ITV}
    \end{minipage}%
    \begin{minipage}[b]{0.24\textwidth}
        \centering
        \includegraphics[width=\textwidth]{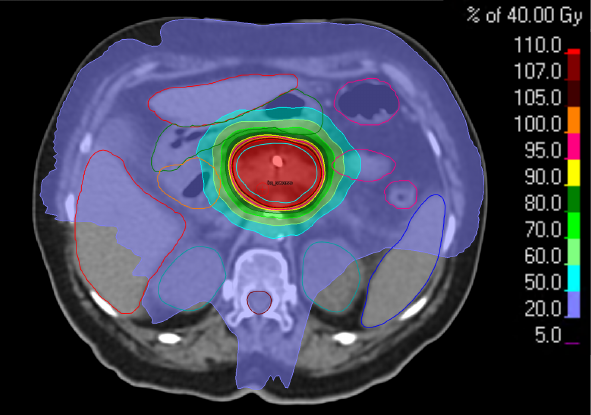}
        \subcaption{SABR 5.5\,cm ITV}
    \end{minipage}
    
    \vskip 0.5cm
    
    \begin{minipage}[b]{0.24\textwidth}
        \centering
        \includegraphics[width=\textwidth]{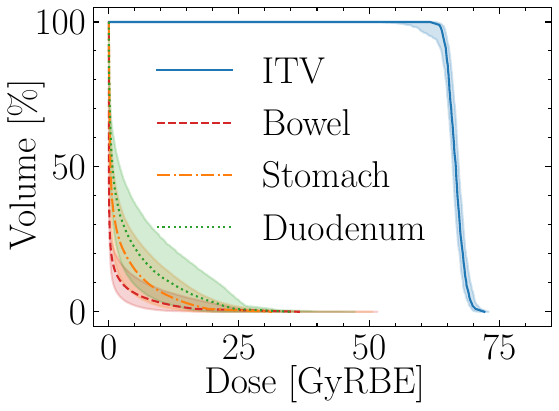}
        \subcaption{dPBT 3.5\,cm ITV}
    \end{minipage}%
    \begin{minipage}[b]{0.24\textwidth}
        \centering
        \includegraphics[width=\textwidth]{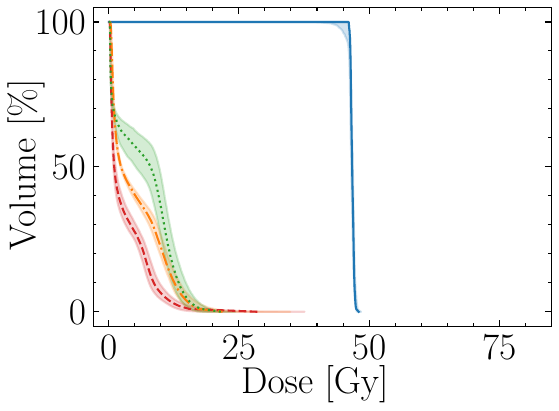}
        \subcaption{SABR 3.5\,cm ITV}
    \end{minipage}%
    \begin{minipage}[b]{0.24\textwidth}
        \centering
        \includegraphics[width=\textwidth]{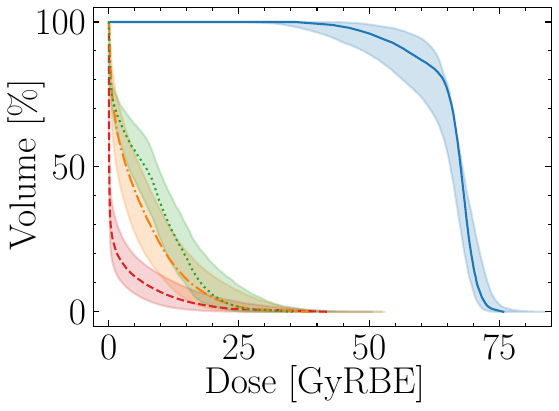}
        \subcaption{dPBT 5.5\,cm ITV}
    \end{minipage}%
    \begin{minipage}[b]{0.24\textwidth}
        \centering
        \includegraphics[width=\textwidth]{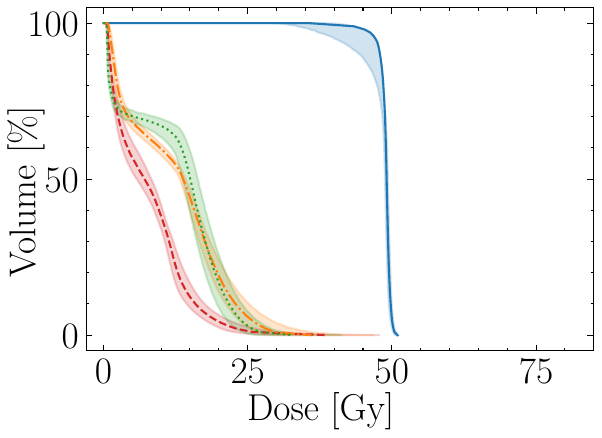}
        \subcaption{SABR 5.5\,cm ITV}
    \end{minipage}  
    \caption{Dose distributions and corresponding DVHs of a dPBT and SABR plan for a 3.5\,cm and 5.5\,cm tumour. Shaded region in each DVH indicates the variation due to anatomical shift scenarios.}
    \label{fig:doseDistDVH}
\end{figure}

 
To quantify the degree to which dose-escalation was achieved with dPBT, we consider $D_{95\%}$ as a function of ITV size in Table \ref{tab:95mets}. Additionally, dPBT and SABR plans for small and large target treatment scenario are shown for each modality in Figure \ref{fig:doseDistDVH}.\par
\begin{figure}[htbp]
   \begin{center}
   \includegraphics[width=10cm]{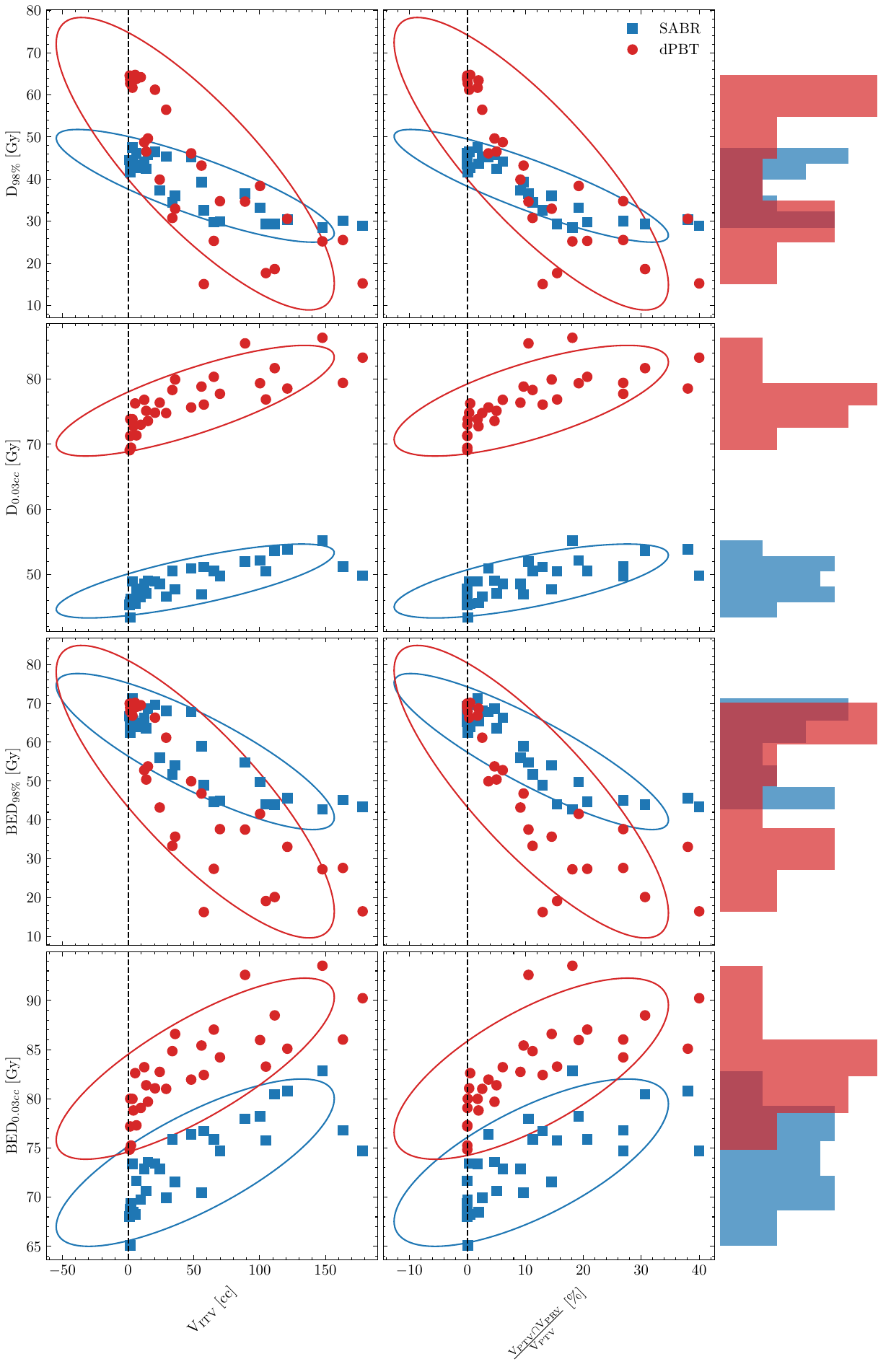}
   \caption{Scatter plot of metrics $D_{98\%}$, $D_{0.03cc}$, $BED_{98\%}$ and $BED_{0.03cc}$ for SABR (blue) and proton (red) as a function of ITV volume and percentage of the PTV overlapping with GI organ PRVs. The error ellipses define 2 standard deviations.
   \label{fig:doseMetricScatterPlot} 
    }  
    \end{center}
\end{figure}

 
As shown in Table \ref{tab:95mets}, dose-escalation to the entire ITV is achievable for sizes up to 5.5\,cm in the nominal scenario. However, in the worst-case scenario, the statistical significance of D$_{95\%}$ diminishes in ITVs larger than 3.5\,cm, suggesting dose-escalation was achieved to $<95$\% of the volume. Notably, dose-escalation is achieved in the portion of the ITV that does not overlap with any gastrointestinal (GI) organs in both nominal and worst-case scenarios, with coverage never falling below that of SABR. Thus suggesting dose-escalation is only feasible for ITV segments that do not overlap with or fall within proximity of a GI structure, making its feasibility inherently patient-dependent, as shown by the large standard deviations in ITV doses for the dPBT plans. Figure \ref{fig:doseMetricScatterPlot} depicts the trend of near-minimum dose (D$_{98\%}$) as a function of V$_{\text{ITV}}$ and 100$\times$(V$_{\text{PTV}}\cap$V$_{\text{PRV}}$)/V$_{\text{PTV}}$. It further supports the notion that patient anatomy and target size remain a major limiting factor for dose-escalation. Furthermore, in an effort to counteract the loss of coverage, the D$_{\text{max}}$ (D$_{\text{0.03\,cm}^3}$) increases with increasing OAR proximity to the ITV for both modalities, although it is more pronounced in the dPBT plans. This loss of uniformity is also depicted by comparing the dose distributions for the SABR and dPBT of a large and small ITV scenario in Figure \ref{fig:doseDistDVH}.\par
The mean (standard deviation) $D_{95\%}$(dPBT) for the smallest target sizes was 64.1(0.7)\,GyRBE in the nominal scenario and 57.3(7.3)\,GyRBE in the worst case, significantly higher ($p$$<$0.05) than $D_{95\%}$(SABR) at 44.4(1.9)\,Gy and 42.1(0.7)\,Gy. In the non-overlapping ITV, we achieve significant dose-escalation across all ITV sizes, however this increase becomes minimal compared to SABR at $\geq$6.5\,cm. Figure \ref{fig:doseDistDVH} depicts a small target (3.5\,cm) treatment scenario and one where coverage was compromised due to the ITV size (5.5\,cm). As arcs are used in the SABR plans, resulting in more degrees of freedom compared to the 2-3 beam dPBT plans, coverage was more uniform and robust compared to dPBT. This is shown by the robust scenario margins between the modalities in Figure \ref{fig:doseDistDVH}). This trend was observed across all treatment scenarios, as shown in the larger difference between the nominal and worst case scenario of $D_{95\%}$(dPBT) compared to $D_{95\%}$(SABR).\par 
\begin{table}[htbp]\centering\caption{Mean (standard deviation) $D_{95\%}$ metric for the whole ITV and the portion of the ITV that does not overlap with any OARs for dPBT (bold) and SABR plans in the nominal (top row) and worst case (bottom row) robustness scenarios. Results of a Wilcoxen signed-rank test for each pair of values for dPBT and SABR in the nominal and worse case scenario are shown in colour. The darkest, lighter and white shaded cells indicate a p-value of $<0.01$, $<0.05$ and $\geq0.05$.}\label{tab:95mets}\begin{tabular}{|c|c|c|c|c|c|c|}\hline
\multirow{2}{*}{\textbf{ROI}}& \multicolumn{6}{c|}{\textbf{Target Size (cm)}} \\
\cline{2-7}  & 2.5 & 3.5 & 4.5 & 5.5 & 6.5 & 7.5 \\
\hline
\multirow{4}{*}{ITV} & \cellcolor{blue!50}\textcolor{black}{\textbf{64.2 (0.7)}} & \cellcolor{blue!50}\textcolor{black}{\textbf{59.8 (7.1)}} & \cellcolor{blue!25}\textbf{51.9 (10.6)} & \cellcolor{blue!25}\textbf{42.3 (8.3)} & \textbf{31.9 (7.4)} & \textbf{26.6 (4.9)} \\
& \cellcolor{blue!50}\textcolor{black}{44.8 (2.0)} & \cellcolor{blue!50}\textcolor{black}{44.3 (1.0)} & \cellcolor{blue!25}42.8 (5.0) & \cellcolor{blue!25}39.1 (6.3) & 33.8 (3.9) & 31.0 (1.2) \\
\cline{2-7}& \cellcolor{blue!25}\textbf{55.5 (7.8)} & \cellcolor{blue!25}\textbf{47.5 (11.4)} & \textbf{34.7 (12.7)} & \textbf{25.0 (7.9)} & \textbf{20.0 (10.4)} & \cellcolor{blue!25}\textbf{16.1 (7.6)} \\
& \cellcolor{blue!25}42.2 (0.6) & \cellcolor{blue!25}38.6 (4.8) & 35.9 (5.2) & 30.5 (2.7) & 29.3 (1.2) & \cellcolor{blue!25}25.7 (2.0) \\
\hline
\multirow{4}{*}{\shortstack{Non-\\overlapping\\ITV}} & \cellcolor{blue!50}\textcolor{black}{\textbf{64.1 (0.8)}} & \cellcolor{blue!50}\textcolor{black}{\textbf{61.0 (5.7)}} & \cellcolor{blue!50}\textcolor{black}{\textbf{56.1 (5.7)}} & \cellcolor{blue!50}\textcolor{black}{\textbf{50.7 (3.2)}} & \cellcolor{blue!25}\textbf{46.4 (4.4)} & \cellcolor{blue!25}\textbf{44.2 (4.5)} \\
& \cellcolor{blue!50}\textcolor{black}{44.8 (2.0)} & \cellcolor{blue!50}\textcolor{black}{44.9 (1.2)} & \cellcolor{blue!50}\textcolor{black}{45.2 (1.5)} & \cellcolor{blue!50}\textcolor{black}{44.0 (2.0)} & \cellcolor{blue!25}42.5 (2.0) & \cellcolor{blue!25}40.1 (2.4) \\
\cline{2-7}& \cellcolor{blue!25}\textbf{55.3 (8.4)} & \cellcolor{blue!25}\textbf{49.2 (9.4)} & \textbf{39.6 (7.6)} & \textbf{33.8 (4.0)} & \textbf{32.1 (3.0)} & \textbf{32.4 (2.0)} \\
& \cellcolor{blue!25}42.2 (0.5) & \cellcolor{blue!25}39.6 (3.8) & 37.2 (4.3) & 32.8 (1.8) & 32.6 (1.2) & 32.0 (1.2) \\
\hline
\end{tabular}\end{table}

\subsection{Factors affecting probability of tumour control}
\begin{figure}[htbp]
    \centering
    \begin{minipage}{0.6\textwidth} 
        \centering
        \begin{subfigure}[b]{\textwidth}
            \centering
            \includegraphics[width=\textwidth]{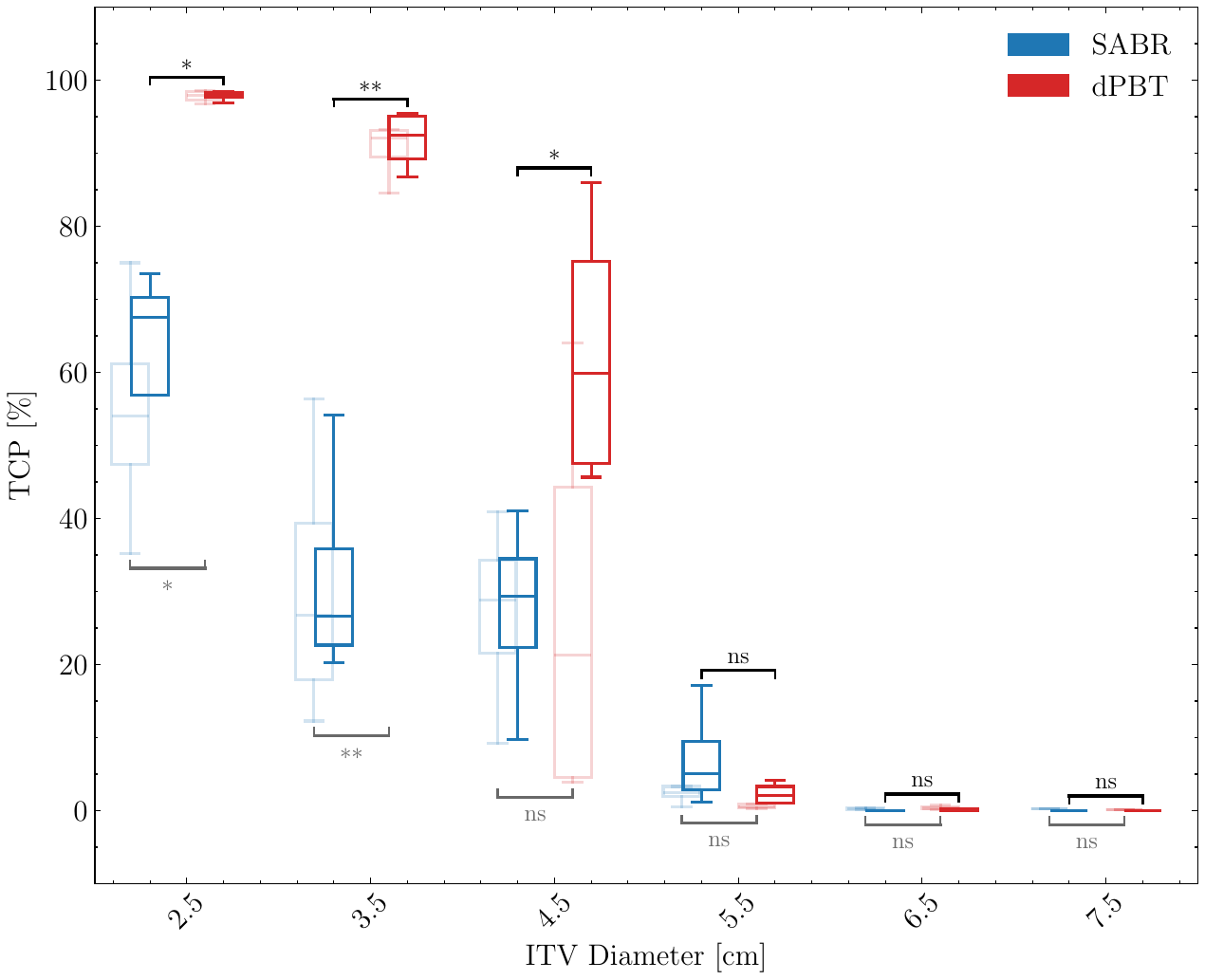} 
            \caption{TCP ranges for each modality.}
            \label{fig:large}
        \end{subfigure}
    \end{minipage}%
    \hfill
    \begin{minipage}{0.4\textwidth} 
        \centering
        \begin{subfigure}[b]{\textwidth}
            \centering
            \includegraphics[width=\textwidth]{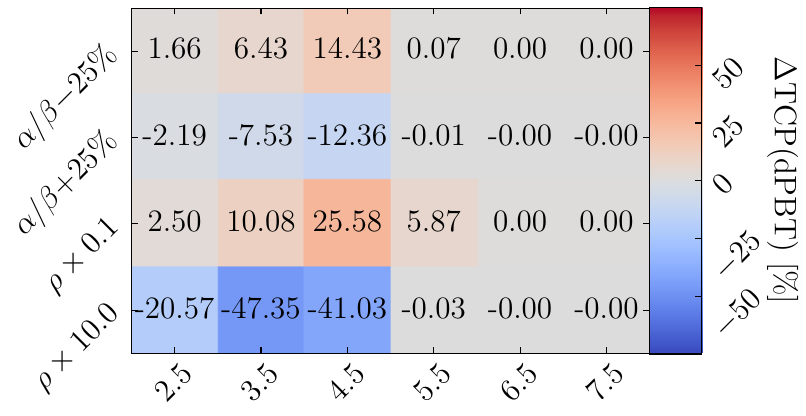} 
            \caption{dPBT parameter sensitivity.}
            \label{fig:small1}
        \end{subfigure}
        \vfill
        \begin{subfigure}[b]{\textwidth}
            \centering
            \includegraphics[width=\textwidth]{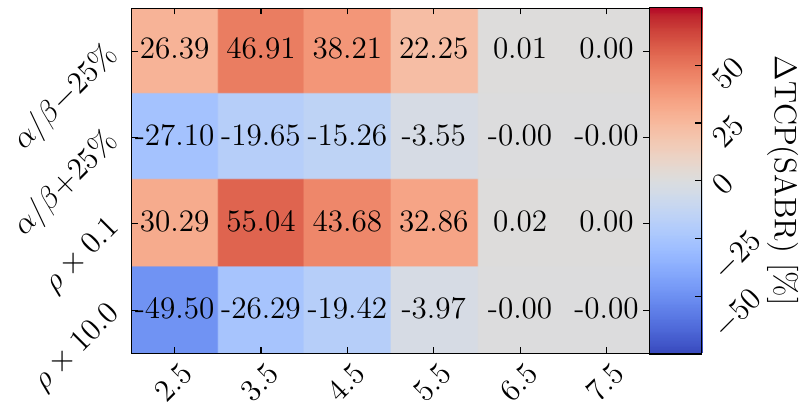} 
            \caption{SABR parameter sensitivity.}
            \label{fig:small2}
        \end{subfigure}
    \end{minipage}
    \caption{TCP predictions (a) for nominal (dark boxes) and worst case (faded boxes) robust scenario. The paired t-test results are indicated for nominal (top) and worst case (bottom) scenarios where $\ast$: $p<$0.05, $\ast\ast$: $p<$0.01, $\ast\ast\ast$: $p<$0.001 and ns: not significant. $\Delta TCP$ [(b), (c)] as a result of varying $\alpha$/$\beta$ by $\pm$25\% and $\rho$ by $\pm$ one order of magnitude.}
    \label{fig:tcpBoxPlot}
\end{figure}

To assess therapeutic benefit, nominal and worst case TCP was evaluated for the ITV in each treatment scenario across the two modalities. By considering the DVH of the ITV across all robust scenarios for each modality, we obtain a range of plausible TCPs when the ITV and OARs are in their average and worst case scenario positions. Figure \ref{fig:tcpBoxPlot} summarises the range of TCPs in the nominal and worst case scenario for different ITV sizes. TCP(dPBT) is $>$95\% and $>$85\% for ITVs with diameter 2.5 and 3.5\,cm, respectively, in all robust scenarios. Whilst the corresponding TCP(SABR), ranged from 35-80\% and 10-60\%. The approximate median target size reported in the literature \cite{fortner1996_lapctargetsizes,comito2017_doseEsc,sun2023_lapctargetsizes} of $\approx$4.5\,cm appears to be a cross-over point where TCP(dPBT) in the nominal scenario ranges from 45-90\%, whilst the worst case TCP(SABR) is comparable with the nominal scenario. This region also appears to be the most uncertain with the widest range due to the sigmoidal nature of the TCP relation. Further, the model parameters derived from the literature to make these predictions also have uncertainty. A subgroup analysis was also conducted to examine TCP in relation to ITV location in the pancreatic head or tail. No correlation with TCP was identified, as increasing ITV size resulted in greater proximity to at least one dose-limiting structure (as defined in Methods). Consequently, large ITVs, whether in the head or tail of the pancreas, were associated with the poorest prognosis due to their higher likelihood of abutting a critical OAR(s).\par
For target sizes exceeding the LAPC median, neither modality saw a therapeutic benefit. This can be explained by the loss in BED, for both modalities in large volume ITVs in Figure \ref{fig:doseMetricScatterPlot}. Whilst dose-escalation was achieved to a portion of these larger ITVs, the BED loss to the whole ITV is detrimental to the TCP. Therefore, we have observed a therapeutic benefit in targets comparable to or smaller than the median LAPC tumour size, however organ proximity is a major limiting factor due to the much lower dose constraints on nearby critical structures.\par

\subsection{Evaluating the safety of dose-escalation}
\begin{figure}[ht]
   \begin{center}
   \includegraphics[width=15cm]{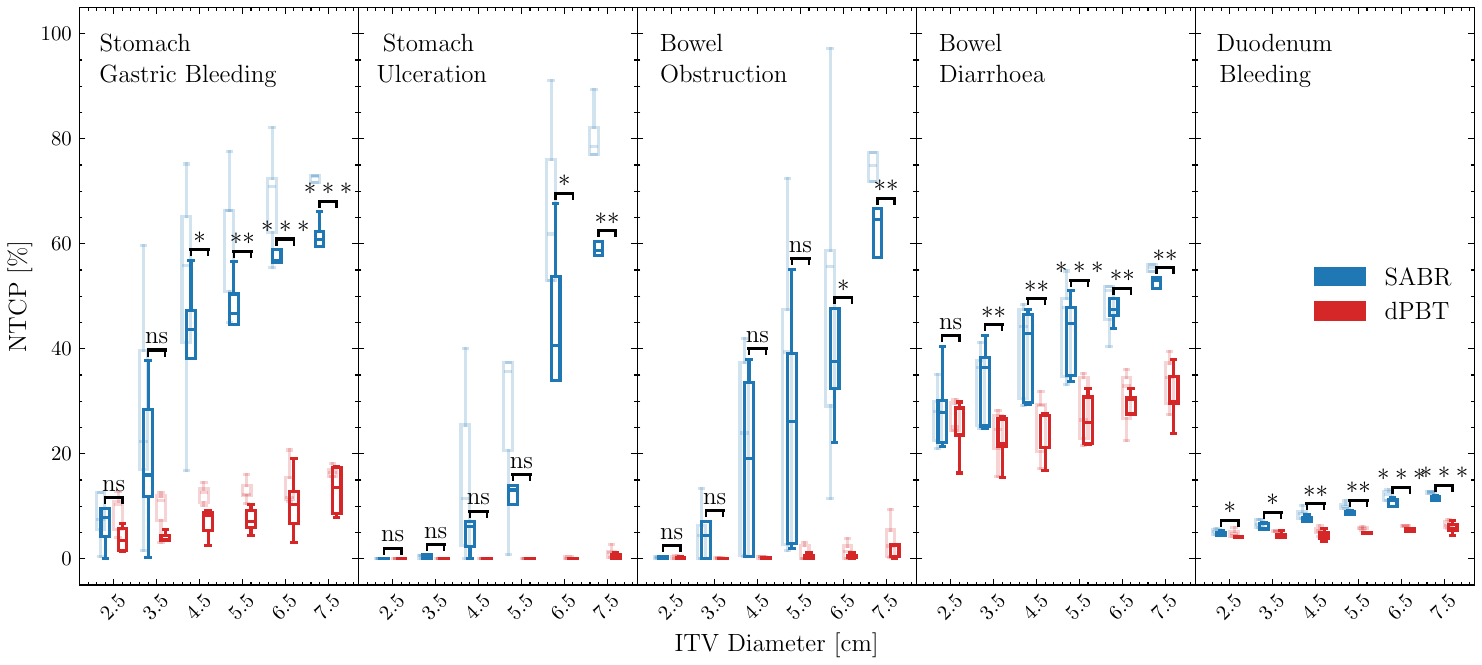}
   \caption{Box plot of NTCPs as a function of ITV size for toxicities in pancreas-proximal GI organs. The darker boxes indicate the nominal treatment scenario, whilst the faded boxes are the worst case scenario. The stars indicate the result of the paired t-test, as indicated in Figure \ref{fig:tcpBoxPlot}.
   \label{fig:ntcpBoxPlot} 
    }  
    \end{center}
\end{figure}

Of primary concern in radiation therapy is limiting the risk of patient toxicities. The NTCP for a range of high and low grade toxicities (Table \ref{tab:lkb_params}) as a function of ITV size is shown in Figure \ref{fig:ntcpBoxPlot}. All GI organs proximal to the pancreas, namely the stomach, bowel and duodenum, have a serial architecture ($n_{\text{NTCP}}\rightarrow0$, Table \ref{tab:lkb_params}) and are therefore more sensitive to maximum dose than integral dose. Figure \ref{fig:doseMetricScatterPlot} shows an increase in D$_{\text{max}}$ with ITV volume and hence V$_{PTV}\cap$V$_{PRV}$. Although D$_{\text{max}}$(dPBT) was consistently higher than D$_{\text{max}}$(SABR), particularly for large V$_{\text{ITV}}$, NTCP(SABR) is significantly higher or comparable to NTCP(dPBT). The reason for this is the difference in fractionation regimens between SABR (8\,Gy/FX) and dPBT (3\,GyRBE/FX).\par
Statistically significant increases of NTCP were mostly seen for bleeding of the stomach with larger targets, as well as low grade toxicities (diarrhoea) regardless of ITV size. As no therapeutic benefit was observed for either modality in targets larger than 4.5\,cm, the NTCPs in this region can be disregarded as these patients are not treatable with radiation therapy and require alternative treatments.\par
For ITV sizes in which a therapeutic benefit was observed (2.5-4.5\,cm), dPBT stomach bleeding probabilities were $<$10\%, compared to that of SABR with up to 8, 20 and 60\% for 2.5, 3.5 and 4.5\,cm ITVs in the nominal scenario, respectively. Stomach ulceration probabilities were near 0\% for dPBT across all targets that saw a therapeutic benefit to dPBT compared to $<$10\% for SABR. A similar trend is observed for bowel obstruction/perforation. The probability of diarrhoea between the two modalities was much more comparable. Although considered a low grade toxicity and potentially compromising patient comfort post-treatment, this toxicity is manageable when weighed against the benefits of increased BED and hence TCP. Due to the comparably small probability of high and low grade toxicities associated with dPBT, it is possible for dose-escalation to be safely achieved using protons despite daily variations in the positions of each region of interest. The aforementioned sub-group analysis did not identify any strong correlation between the NTCP for the select toxicities and OARs.\par

\subsection{Parameter sensitivity of TCP and NTCP predictions}
Figures \ref{fig:small1} and \ref{fig:small2} depict the change in mean TCP when $\alpha/\beta$ and $\rho$ are varied independently for the dPBT and SABR plans, respectively. Overall, more parameter sensitivity is observed for TCP(SABR) compared to TCP(dPBT) due to TCP(dPBT) being generally closer to the flat portion of the sigmoidal TCP relation. Although, the effect of $\alpha/\beta$ uncertainty can change TCP(dPBT) up to 14.4\% and TCP(SABR) up to 46.9\%. The more influential parameter for both modalities is $\rho$, with changes up to 55\% and 47.3\% in TCP(SABR) and TCP(dPBT), respectively. We chose $\rho$ to be a conservative number of 10$^8$ clonogens/cm$^3$ for our predictions for this reason. Overall, we note that both parameters are highly influential to the TCP predictions made in this work.\par
Depicted in Figures \ref{fig:ntcpParameterSensitivityPlotdPBT} and \ref{fig:ntcpParameterSensitivityPlotSABR} are the mean variations of NTCP(dPBT) and NTCP(SABR), respectively, for the toxicities considered in Figure \ref{fig:ntcpBoxPlot} when the LKB model parameters from Table \ref{tab:lkb_params} are varied by $\pm25\%$. NTCP predictions were found to be most sensitive to variations in TD$_{50}$, whilst variations in $m$ and $n$ resulted in only minor NTCP deviations. For target sizes 2.5-4.5\,cm, a 25\% decrease in TD$_{50}$ the mean NTCP(SABR) was up to 45\% higher than the predictions in Figure \ref{fig:ntcpBoxPlot}, whilst NTCP(dPBT) for the same target size range only increased up to 33\%. For larger target sizes, the NTCP predictions vary more than the smaller targets, however as there is no therapeutic benefit demonstrated by the low TCPs in Figure \ref{fig:tcpBoxPlot} for targets 5.5\,cm and larger, the NTCPs can be disregarded as our results suggest dPBT is infeasible for these cases. The NTCP predictions were also observed to be sensitive to variations in $\alpha/\beta$. In this work an $\alpha/\beta$ of 3 was assumed for all OARs. For target sizes 2.5-4.5\,cm, a 25\% decrease in $\alpha/\beta$ resulted in NTCPs increasing by up to 10\%, whilst a 25\% increase in $\alpha/\beta$ resulted in a NTCP decrease as much as 6.7\%. Finally, variations of LKB parameters $m$ and $n$ saw only minor changes to NTCP predictions of up to 5\% for both modalities. Overall, NTCP(dPBT) remains either comparable to or lower than NTCP(SABR) in light of uncertainties in the radiobiological parameters summarised in Table \ref{tab:lkb_params}.\par

\begin{table}[ht]
   \begin{center}
   \includegraphics[width=15cm]{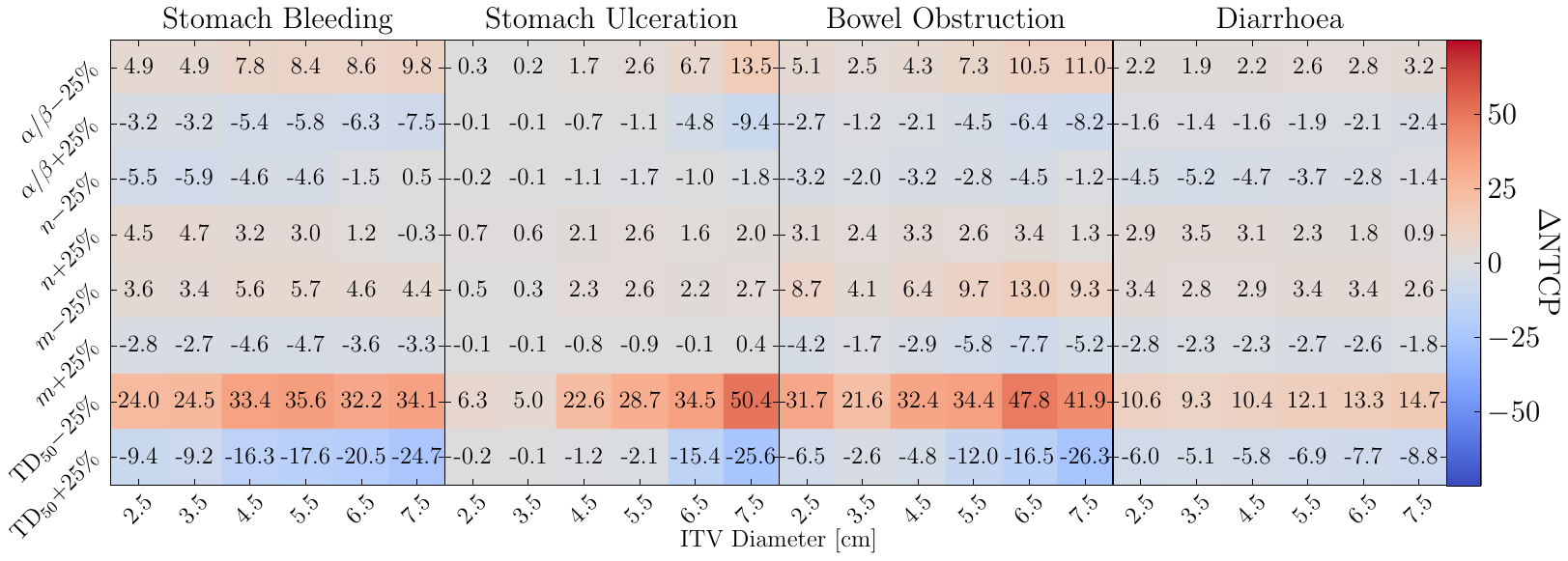}
   \caption{Sensitivity of NTCP calculations for select toxicities for dPBT plans when parameters from the Lyman-Kutcher-Burman Model, $m$, $n$ and TD$_{50}\pm$25\%.
   \label{fig:ntcpParameterSensitivityPlotdPBT} 
    }  
    \end{center}
\end{table}

\begin{table}[ht]
   \begin{center}
   \includegraphics[width=15cm]{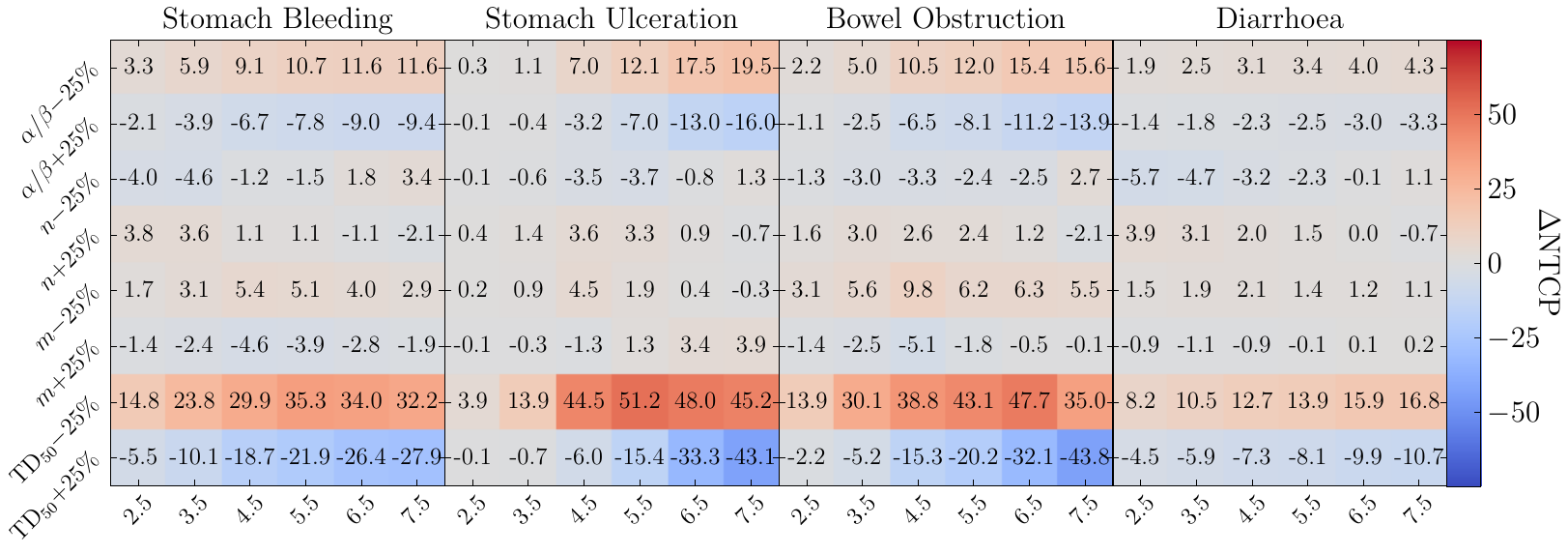}
   \caption{Sensitivity of NTCP calculations for select toxicities for SABR plans when parameters from the Lyman-Kutcher-Burman Model, $m$, $n$ and TD$_{50}\pm$25\%.
   \label{fig:ntcpParameterSensitivityPlotSABR} 
    }  
    \end{center}
\end{table}

\subsection{The effect of fraction regimen}
The significantly lower NTCPs resultant of the dPBT fractionation regimen, compared to SABR, suggests either the dose can be boosted further, or the current dPBT prescription can be hypofractionated further. The fractionation regimen was varied until the NTCP(dPBT)s considered in Figure \ref{fig:ntcpBoxPlot} were similar to, or no worse than SABR. This limit was found to be 66\,GyRBE in 10 fractions, as opposed to 22 fractions in the original protocol.\par
\begin{figure}[ht]
   \begin{center}
   \includegraphics[width=15cm]{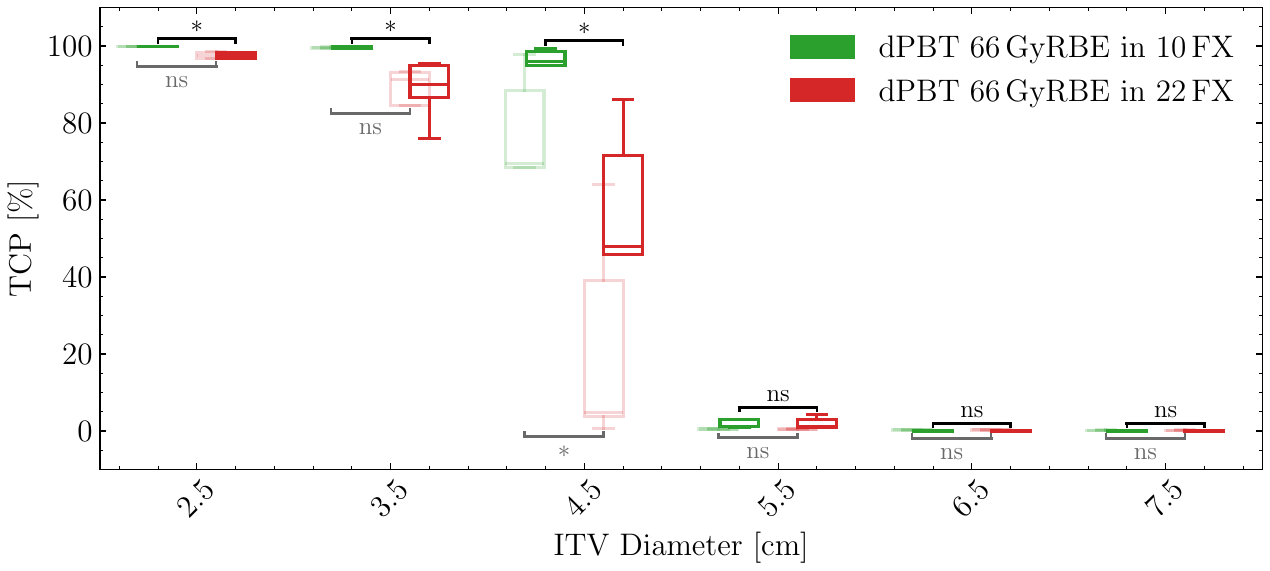}
   \caption{Box plot of TCPs, where the dPBT fraction regimen is 66\,GyRBE in 10 fractions, as a function of ITV size. The darker boxes indicate the nominal treatment scenario, whilst the faded boxes are the worst case scenario. The stars indicate the results of the paired t-test.
   \label{fig:tcpBoxPlot_10FX} 
    }  
    \end{center}
\end{figure}

As shown in Figure \ref{fig:tcpBoxPlot_10FX}, an alternative fractionation regimen improved the TCP in all uncertainty scenarios for ITV diameters $<$4.5\,cm by a small amount. For targets 4.5\,cm in size, the therapeutic benefit was notably improved compared to the original schedule, particularly the worst case scenario, which is now comparable to the nominal TCP. However, there remains no therapeutic benefit to hypofractionation for ITVs larger than 5.5\,cm, as the TCP did not increase from 0\% when the dose per fraction was increased to 6.6\,GyRBE.

\section{Discussion}
\subsection{Dosimetry \& limitations of GI organ proximity}
In this study, we evaluated the feasibility of dose-escalation to the pancreas using PBT for 30 treatment scenarios, each with large variation in GI organ and target position, as well as its size and intra-tumour radiosensitivity. We determined a range of feasible TCPs and probabilities for select high and low grade toxicities, and their variations increasing ITV size. The extent to which dose-escalation was achieved in each scenario using PBT informed any change in therapeutic benefit and resultant toxicities compared to SABR.\par 
We found that dose-escalation is inherently patient-dependent and limited by abutting GI organs, such as the stomach, duodenum and bowel, as they are assigned a higher priority than target coverage - a finding echoed in the literature \cite{Bertholet2019_doseEsc, Crane2016_doseEsc, krishnan2015_doseEsc}. Generally, dose-escalation is achievable using dPBT in the portion of the ITV that does not overlap with any GI organ when the anatomy is in its nominal position, as demonstrated by the statistically significant increase of D$_{95\%}$(dPBT) compared to that of SABR shown in Table \ref{tab:95mets}. However, for larger targets where overlap is observed in the nominal case as well as the robust scenarios, the deliverable BED to the ITV diminishes below that of SABR. BED has previously been reported as a major predictor of tumour control and overall survival \cite{krishnan2015_doseEsc, Bertholet2019_doseEsc}, and as such, a multimodality, patient-specific approach is needed to either boost the sensitivity of these OAR-proximal areas and may improve tumour control. This may be achievable with a combined dPBT and chemotherapy treatment \cite{krishnan2015_doseEsc,comito2017_doseEsc} in larger ITVs positioned to GI structures at a distance within the robust uncertainties.\par
Increased degrees of freedom through the use of arcs for SABR allowed for better dose shaping than dPBT, which used up to three discrete static beams. Similar sentiments have been made in previous dosimetric studies of PBT for LAPC \cite{bouchard2009_doseEsc}. This can possibly be explained by an elevated D$_{\text{max}}$(dPBT) compared to D$_{\text{max}}$(SABR). It suggests more beams are needed to ensure hotspots are not present in vicinity of serial GI organs - this is a significant limiting factor for dPBT, as vast improvements are observed in integral dose to GI structures with dPBT, whereas BED$_{\text{max}}$(dPBT) is only slightly elevated compared to BED$_{\text{max}}$(SABR). One possible solution to this is the use of proton arc therapy \cite{Ding2016_sparc}; although arcs would be limited to the recommended posterior-anterior direction \cite{AMOS2022e188}.\par 
\subsection{Therapeutic benefit \& volume dependence}
The therapeutic benefit of dPBT was limited to targets $\leq$4.5\,cm compared to SABR in the nominal treatment scenario. Whilst in the worst case scenario, the TCP at 4.5\,cm was comparable to both the nominal and worst case TCP of SABR. Thereafter, neither modality saw a therapeutic benefit. Within the treatment scenarios considered in this work, targets with a diameter of 4.5\,cm may require a patient-specific approach to dose-escalation throughout treatment. The worst case scenario limits the obtainable BED inside the ITV for larger volumes, however this scenario may not occur frequently. As LAPC patients are imaged prior to the delivery of each treatment fraction \cite{masterplan}, a selective BED boost approach may be taken when anatomy is favourably positioned, and additional precautions are taken when it is not.\par 

\subsection{dPBT safety \& toxicities}
We defined dPBT to be safe if the NTCP of select toxicities was similar to or less than that of SABR. Due to the sensitivity of OARs to hypofractionation, the NTCPs of SABR was consistently higher than dPBT. Thus at the fractionation used in the dPBT protocol, dose-escalation can be safely achieved. As the most proximal OARs to the ITV are serial, they are sensitive to maximum dose as opposed to integral dose. As such, the dPBT plans may benefit from linear energy transfer (LET) optimisation to minimise the enhanced biological effect of LET hotspots inside these structures \cite{McIntyre2023_LETopt}.\par

\subsection{The role of hypofractionation}
In order to boost the BED of dPBT, we increased the dose per fraction from the regimen used in the protocol until NTCP(dPBT) for the toxicities considered above were comparable to or no worst than NTCP(SABR) in the worst case scenario. For our dPBT plans, this limit was 66\,GyRBE in 10 fractions, as opposed to 22 fractions in the original protocol. This results in a prescription BED boost from 86 to 109.6\,GyRBE (assuming $\alpha$/$\beta$=10), leading to significant improvements to TCP(dPBT) for ITVs with a diameter of 4.5\,cm in both nominal and worst case scenarios. LAPC incidence data suggests that up to 70\% of tumours are located in the head of the pancreas, whilst the remainder occur in the pancreas body and tail \cite{fortner1996_lapctargetsizes, sun2023_lapctargetsizes}. LAPC of the pancreas tail is often diagnosed later than that of the head, and as such, tumours are often in the order of 6\,cm in the tail compared to 3\,cm in the head \cite{Vareedayah2018_lapcstats}. This suggests that a therapeutic benefit for smaller targets may be promising for higher incidence LAPC characteristics.\par 

\subsection{Limitations \& future work}
There are a number of limitations that may influence the results of this study. It is noted that the effects of hypoxia were omitted in the current work. As a result, the TCPs are likely overestimated for both modalities. Due to the 4.5\,cm target size region having the widest margin of uncertainty, depicted in Figure \ref{fig:tcpBoxPlot}, this is where the largest discrepancy is expected. Further, the input parameters for both the TCP and NTCP models were derived from clinical data based on photon treatments and were used for our dPBT calculations due to a lack of equivalent proton data. A dependence of LET on $\alpha/\beta$ has been previously observed in human cell lines \cite{Mara2020_cellsurvivalstudy} and suggests that the equivalent proton tumour $\alpha/\beta$ may be lower. Our TCP parameter sensitivity analysis suggests that a 25\% decrease in $\alpha/\beta$ will see an increase in TCP(dPBT), particularly for ITVs with a 5.5\,cm diameter (Figure \ref{fig:small1}).\par
The impact of intra-fraction anatomical changes has suggested the need for online-adaptive planning techniques using daily CBCT images \cite{cbctdaily}. In the protocols utilised in this study, daily CBCTs were taken of each patient prior to the delivery of each fraction in addition to strict pre-treatment preparation measures such as fasting and the delivery of each fraction at the same time of day. In our study, these shifts were assumed to lie within the 5\,mm margin considered, which we acknowledge as a limitation. With an online-adaptive approach, the TCP could increase or decrease according to the daily shifts, particularly for a median ITV size.\par
Despite these limitations, the overarching conclusion does not change in that OAR proximity is the primary factor that limits not only dPBT efficacy, but dose-escalated radiotherapy in general. Whilst changes to the computed TCPs above are especially anticipated as a result of these limitations for patients with a median-sized ITV, it still suggests that a patient-specific, case-by-case approach is required for these patients to determine whether dPBT or other dose-escalated radiotherapy treatments can be safely delivered with a favourable outcome.\par
The effects of tumour hypoxia will be subject to further study. The effect on TCP will be quantified, with and without hypoxic sensitising drugs. Further, the role of LET optimisation for LAPC will be explored to maximize the biological effect of PBT. Its impact is still an active area of investigation, however the literature thus far has found cases that use asymmetric beam arrangements can result in enhanced LET hot spots due to the concentration of high LET proton track ends occurring in the distal region of the Bragg Peak \cite{McIntyre2023_LETopt} to a small area. As beam arrangement is strictly limited by surrounding anatomy, shifting LET hot spots outside the proximal GI structures may further improve the safety of dPBT. 

\section{Conclusion}
Across the simulated patient cohort, clinical scenarios, and uncertainties considered in this work,
\begin{itemize}
    \item dose escalation using proton beam therapy can be safely achieved for patients with target sizes up to the median reported in the literature, resulting in an improvement to local tumour control.
    \item the close proximity of the stomach, duodenum and bowel remains a major limiting factor for effective radiation treatments of LAPC.
    \item patient-specific outcome modeling provides valuable insight into the likely efficacy of radiation treatment and this patient-specific approach may be beneficial through adaptive planning methods.
    \item an extended patient cohort and long-term follow up would be required to conclusively determine which patients would benefit, however the results in the current study are promising for patients with small to median-sized tumours.
    \item increasing the BED through hypofractionation further improves the therapeutic benefit for targets with minor overlap of the stomach, duodenum and bowel, however this can result in increased toxicities comparable to the current standard of radiation care for LAPC.
\end{itemize}

\section*{Acknowledgements}
M. A. M. is supported by the Australian Government Research Training Program scholarship. R.C.-E.H. was supported by grants (NSTC111-2314-B-182A-160-MY2 and NSTC112-2628-B-182A-007-MY3) from the National Science and Technology Council, as well as the Yushan Fellow Program (MOE-113-YSFMN-1009-001-P1) from the Ministry of Education in Taiwan.



\end{document}